# Harmonic-Aware Transformer for Real-Time Catheter Localization in Interventional Procedures of Magnetic Particle Imaging

Abuobaida M. Khair[1,2], Wenjing Jiang[1] Xiaoli Yang[3] Moritz Wildgruber[4] and Xiaopeng Ma[1,*]

[1]School of Control Science and Engineering, Shandong University, Jinan 250061, China
[2]Department of Medical Physics, Al-Neelain University, Khartoum, Sudan
[3]Weifang Xinli Superconducting Magnet Technology Co., Ltd, Weifang 261005, China
[4]Department of Radiology, University Hospital, LMU Munich, Munich 81337, Germany

**E-mail:** xiaopeng.ma@sdu.edu.cn

## Abstract

*Objective.* Magnetic Particle Imaging (MPI) enables real-time, radiation-free tracking of magnetic nanoparticle–coated instruments, making it highly suitable for interventional procedures. This study proposes a harmonic-aware transformer framework that directly predicts catheter tip positions from raw MPI voltage signals, eliminating the need for image reconstruction and reducing computational latency.
*Approach.* The framework incorporates frequency-domain preprocessing to isolate the 2nd–8th drive-field harmonics, enhancing the signal-to-noise ratio (SNR) while preserving motion-relevant features. A transformer architecture with six encoder layers and eight attention heads is employed to learn spatio-temporal dependencies across the three receive axes ($x$, $y$, $z$) for accurate three dimensional position estimation. The model is trained on simulated MPI signals and evaluated on real in vitro datasets under standard, bending, and heartbeat-like motion conditions.
*Main results.* The proposed method achieves sub-millimeter localization accuracy, with a minimum L2 error of 0.103 $\pm$ 0.092 mm and mean absolute errors (MAE) of 0.039 $\pm$ 0.046 mm, 0.054 $\pm$ 0.049 mm, and 0.060 $\pm$ 0.044 mm along the ($x, y, z$) axes, respectively, for the bending dataset. Across all datasets, the MAE ranges from 0.165 mm to 0.655 mm, demonstrating consistent performance. The optimized inference achieves a latency of 0.55 ms per frame and a throughput of approximately 1800 frames/s, confirming real-time capability.
*Significance.* Compared with conventional MPI-guided approaches relying on image reconstruction, the proposed framework provides improved accuracy, reduced latency, and enhanced robustness under complex motion conditions. These results highlight the potential of harmonic-aware transformer models as efficient and scalable solutions for real-time catheter localization in interventional MPI.

## 1 Introduction

Magnetic particle imaging (MPI) is a cutting-edge medical imaging technology that enables the precise visualization of processes within the body. Unlike traditional imaging methods, MPI directly detects magnetic nanoparticles (MNPs), which serve as tracers that can be safely injected and tracked in real-time (Han *et al*, 2020; Rezaei *et al*, 2024).

The technique relies on how these nanoparticles respond to specially designed magnetic fields, one static (selection field) and one oscillating (drive field), generated by coils within the scanner. When the magnetic fields interact with the nanoparticles, they produce unique signals that are picked up by receiving coils. These signals are then used to reconstruct detailed images that accurately depict the tracers' positions and concentrations within the body (Panagiotopoulos *et al*, 2015).

MPI tracks the spatiotemporal distribution of magnetic nanoparticles, producing 4-dimensional images with very high temporal resolution. This is made possible by leveraging the nonlinear magnetization behavior of the nanoparticles, enabling MPI to monitor their movement and concentration changes inside the body quickly and accurately (Gleich and Weizenecker, 2005; Weizenecker *et al*, 2009).

Beyond its use in cell tracking (Zheng *et al*, 2015; Them *et al*, 2016), the high temporal resolution and real-time imaging capability of MPI make it highly promising for interventional applications. Unlike conventional X-ray-based digital subtraction angiography (DSA), which relies on ionizing radiation and contrast agents, MPI offers a radiation-free alternative that is both safe and fast. Its ability to visualize magnetic tracers in real time allows clinicians to guide instruments such as catheters with exceptional precision. This potential has already been demonstrated in previous studies (Haegele *et al*, 2012; Griese *et al*, 2017; Hartung *et al*, 2025; Ludewig *et al*, 2022).

Recent proof-of-concept studies demonstrated MPI's potential for cardiovascular imaging and interventional guidance Weizenecker *et al* (2009); Salamon et al. (2016); Herz *et al* (2018, 2019); Ahlborg et al. (2022). For interventional procedures, it is crucial to accurately determine the position of instruments or markers coated with magnetic tracers Werner *et al* (2016).

In practice, MPI spatial resolution is primarily influenced by nanoparticle properties and the selection field gradient (Griese *et al*, 2017). Traditional MPI reconstruction methods (system-matrix approach or X-space) have limitations for real-time use: the system matrix requires large calibration scans and high memory; X-space can be faster but can produce blur (Knopp and Buzug, 2012; Guo *et al*, 2025).

Many deep learning approaches target image quality and reconstruction speed (Zhao *et al*, 2024; Wen *et al*, 2024; Gapyak *et al*, 2025), but typically rely on image formation as an intermediate step.

Recent advancements in MPI increasingly leverage deep learning to improve signal quality, image clarity, and reconstruction speed. Several studies use neural networks to denoise low signal-to-noise-ratio (SNR) data or accelerate reconstruction, improving temporal resolution without requiring hardware modifications (Shi *et al*, 2025; Sun *et al*, 2025). However, these methods primarily focus on producing enhanced images rather than tracking interventional devices. Early MPI-guided intervention work demonstrated real-time tracking but relied on traditional processing pipelines. Rahmer *et al* (2017) achieved magnetic catheter tracking using multi-color MPI and force control, while Griese *et al* (2017) localized fiducial markers using center-of-mass estimation.Herz *et al* (2018) enabled real-time visualization of labeled guidewires and balloons, and later work extended tracking to Intravascular Optical Coherence Tomography (IVOCT). catheter tips (Griese *et al*, 2020) and MNP-embedded balloons (Ahlborg et al., 2022). These approaches did not incorporate deep learning or harmonic-domain preprocessing.

Other studies focused on reconstruction instead of tracking. Shen *et al* (2023) improved iterative reconstruction on open phantom data, Wen *et al* (2024) proposed a a convolutional neural network (CNN) for point-spread-function (PSF)-aware image deblurring, and Wegner *et al* (2024) visualized unmodified stents using Kaczmarz reconstruction. More recent deep learning work improved image quality through transformer-based frequency fusion (Sun *et al*, 2025) or CNN-based distillation (Shi *et al*, 2025), but still did not address catheter localization.

In contrast, our method directly maps harmonically preprocessed MPI signals to three-dimensional (3D) catheter coordinates using a transformer backbone. This shifts the focus from image reconstruction to real-time catheter tip tracking, leveraging both simulated and real datasets.

Our primary contributions are summarized as follows:

A reconstruction-free, signal-driven framework that directly maps raw MPI voltage signals to catheter tip positions, eliminating the conventional system matrix inversion and image reconstruction pipeline.

A harmonic-aware preprocessing strategy that selectively extracts the 2nd–8th drive-field harmonics to enhance SNR while retaining motion-sensitive spectral components.

A transformer-based architecture specifically designed for MPI signals, enabling effective modeling of temporal dynamics and cross-axis ($x$, $y$, $z$) correlations for accurate 3D localization.

Extensive validation on both simulated and real in vitro MPI datasets, demonstrating robust sub-millimeter accuracy under diverse motion patterns and achieving real-time inference performance.

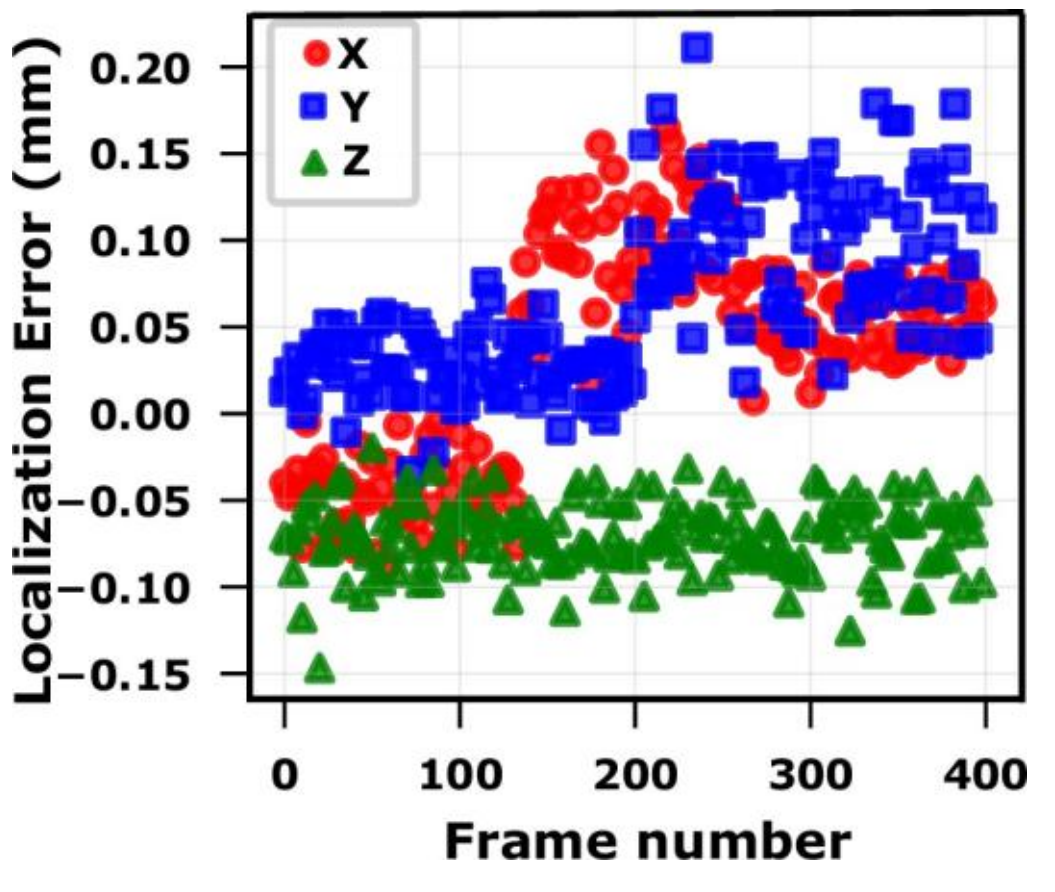

(a) Localization of the catheter position estimated using experimental data.

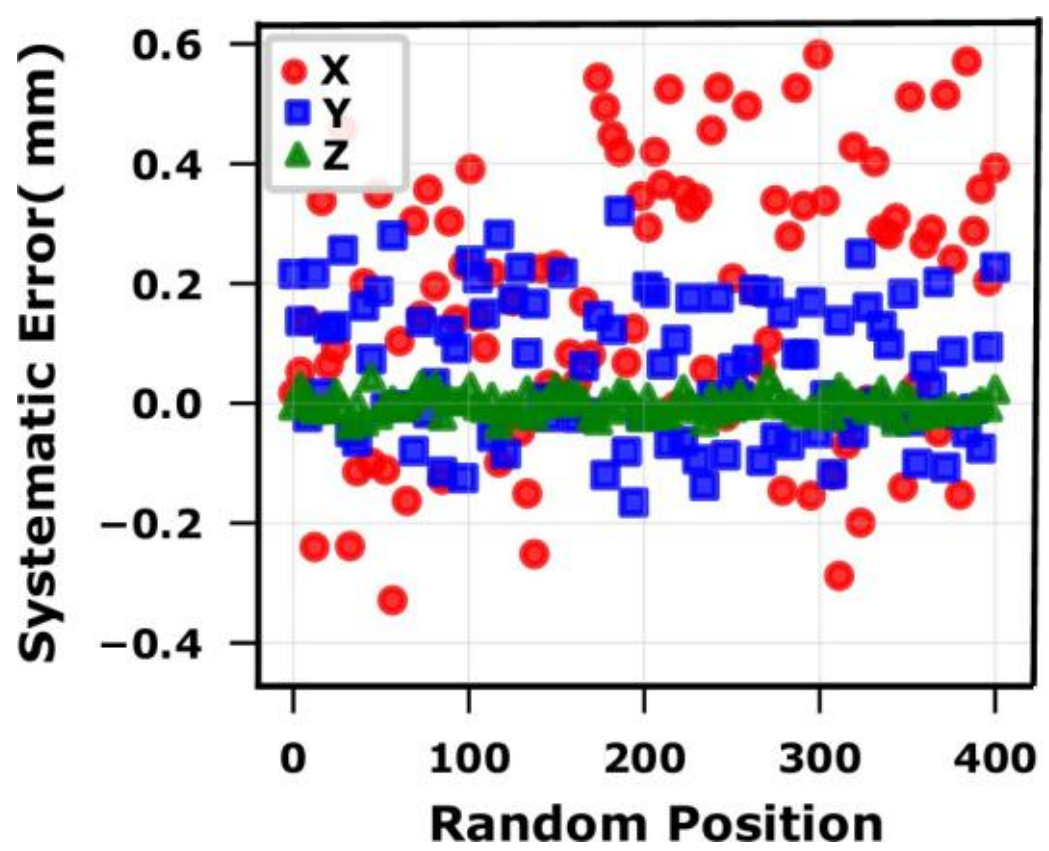

(b) Systematic error in catheter localization estimates derived from phantom experiments.

Figure 1: The catheter position estimated using experimental data as reported in (Griese *et al* , 2017). (a) illustrates the localization, while (b) shows the systematic error in catheter localization estimates derived from measurements taken during phantom experiments.

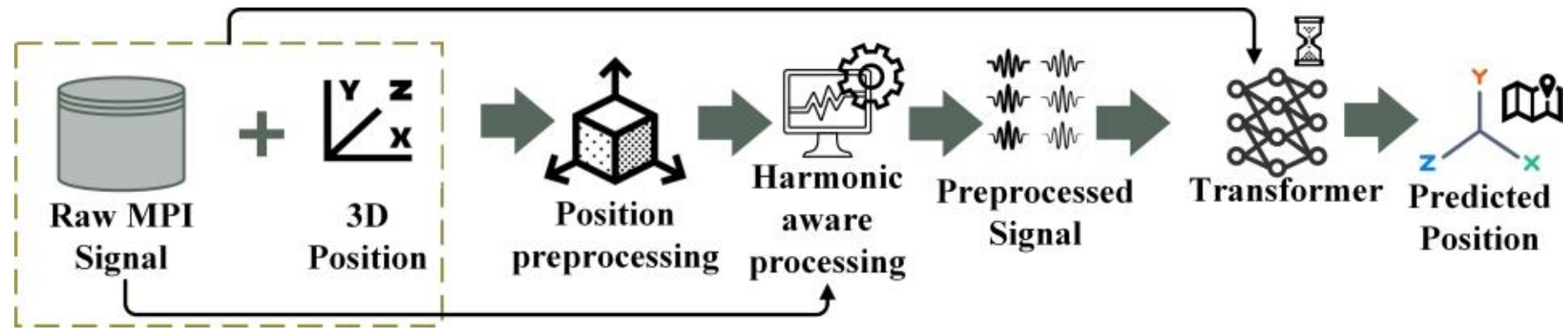

Figure 2: Overview of the proposed method.

## 2 Methods

We propose a harmonic-aware preprocessing pipeline followed by a transformer encoder and a small MLP regression head. The pipeline transforms the complex-valued time domain MPI signals into a compact frequency-domain representation that preserves the harmonics that encode positional information. The transformer then learns spatio–temporal spectral patterns to regress 3D catheter tip coordinates.

### *2.1 Model Architecture*

A transformer-based model, referred to as the catheter transformer, is developed to directly estimate the 3D position of the catheter tip from frequency-domain MPI signals. The transformer architecture is leveraged to learn a nonlinear mapping between complex frequency patterns and corresponding spatial coordinates by capturing long-range dependencies in the data.

Each MPI sample is represented as a one-dimensional signal with three channels:

$$X \in \mathbb{R}^{3\times L} \tag{1}$$

where $L$ is the number of frequency components, and the three channels correspond to the received signals along the *x*-, *y*-, and *z*-directions.

To process these long signals efficiently, the catheter transformer first divides the input into non-overlapping patches using a 1D convolutional layer. This layer performs patch embedding with kernel size and stride $P = 64$, generating compact tokens that summarize local frequency patterns. The resulting embedded sequence is denoted as:

$$Z = \text{Conv1D}(X, \text{kernel} = 64, \text{stride} = 64), \tag{2}$$

where $Z \in \mathbb{R}^{d_{\text{model}} \times N_p}$, $d_{\text{model}}$ is the embedding dimension, and $N_p$ is the number of patches. Each column of $Z$ corresponds to a latent representation of a local spectral segment.

To incorporate positional information, sinusoidal positional encoding (PE) is added to the embedded patches. For position $i$ and dimension $k$, the encoding is defined as:

$$\mathrm{PE}(i, 2k) = \sin\left(\frac{i}{f^{2k/d_{\mathrm{model}}}}\right), \tag{3}$$

$$\mathrm{PE}(i, 2k+1) = \cos\left(\frac{i}{f^{2k/d_{\mathrm{model}}}}\right). \tag{4}$$

The encoded input sequence becomes:

$$\tilde{Z} = Z + \mathrm{PE}. \tag{5}$$

The sequence $\tilde{Z}$ is then processed by six transformer encoder layers. Each layer employs multi-head self-attention (MHSA) with $H = 8$ heads to capture global dependencies across spectral patches. The attention mechanism is defined as:

$$Q = \tilde{Z} W_Q, \quad K = \tilde{Z} W_K, \quad V = \tilde{Z} W_V, \tag{6}$$

where $W_Q, W_K, W_V \in \mathbb{R}^{d_{\mathrm{model}} \times d_h}$ are learned projection matrices, and $d_h = d_{\mathrm{model}}/H$ is the dimension of each attention head.

Self-attention is computed as:

$$A = \mathrm{softmax}\left(\frac{QK^\top}{\sqrt{d_h}}\right) V. \tag{7}$$

The outputs from all heads are concatenated and linearly projected:

$$\mathrm{MHSA}(\tilde{Z}) = \mathrm{Concat}(A_1, \ldots, A_H) W_A. \tag{8}$$

A residual connection followed by layer normalization is applied:

$$Z' = \mathrm{LayerNorm}\left(\tilde{Z} + \mathrm{MHSA}(\tilde{Z})\right). \tag{9}$$

Each encoder layer also includes a position-wise feedforward network (FFN) with GELU activation and hidden dimension $d_f = 512$:

$$\mathrm{FFN}(Z') = W_2\, \mathrm{GELU}(W_1 Z' + b_1) + b_2, \tag{10}$$

where $W_1 \in \mathbb{R}^{d_f \times d_{\mathrm{model}}}$ and $W_2 \in \mathbb{R}^{d_{\mathrm{model}} \times d_f}$.

A second residual connection yields the final encoded representation:

$$Z_{\mathrm{enc}} = \mathrm{LayerNorm}\left(Z' + \mathrm{FFN}(Z')\right). \tag{11}$$

After all encoder layers, the encoded sequence is flattened into a feature vector $Z_f$, which is passed through a two-layer multilayer perceptron (MLP) for coordinate regression:

$$h = \mathrm{GELU}(W_1 Z_f + b_1), \tag{12}$$

$$\hat{\mathbf{y}} = W_2 h + b_2, \tag{13}$$

where $\hat{\mathbf{y}} = [\hat{x}, \hat{y}, \hat{z}]^\top$ denotes the predicted 3D catheter tip position.

The model is trained by minimizing the loss function $\mathcal{L}$, defined as a weighted mean absolute error across all spatial dimensions:

$$\mathcal{L} = \frac{1}{N} \sum_{j=1}^{N} \left( w_x |x_j - \hat{x}_j| + w_y |y_j - \hat{y}_j| + w_z |z_j - \hat{z}_j| \right) \tag{14}$$

where $w_z > w_x = w_y$ to compensate for the lower spatial resolution along the Z-axis in MPI, $N$ denotes the total number of samples in the batch, $(x_j, y_j, z_j)$ and $(\hat{x}_j, \hat{y}_j, \hat{z}_j)$ are the ground-truth and predicted spatial coordinates for the $j$-th sample, respectively, and $w_x$, $w_y$, and $w_z$ are the axis-wise weighting factors. Equation 14 represents the primary training objective used to optimize the proposed transformer model.

Additionally, the $\mathcal{L}2$ error evaluates the Euclidean distance between predicted and true 3D positions:

$$\mathcal{L}2 = \frac{1}{N} \sum_{j=1}^{N} \sqrt{(\hat{y}_{j,x} - y_{j,x})^2 + (\hat{y}_{j,y} - y_{j,y})^2 + (\hat{y}_{j,z} - y_{j,z})^2}. \tag{15}$$

Equation 15 defines the Euclidean (L2) localization error used for performance evaluation after training. Unlike Equation 14, this metric is not used for parameter optimization, but instead measures the overall 3D distance between the predicted and ground-truth catheter tip positions.

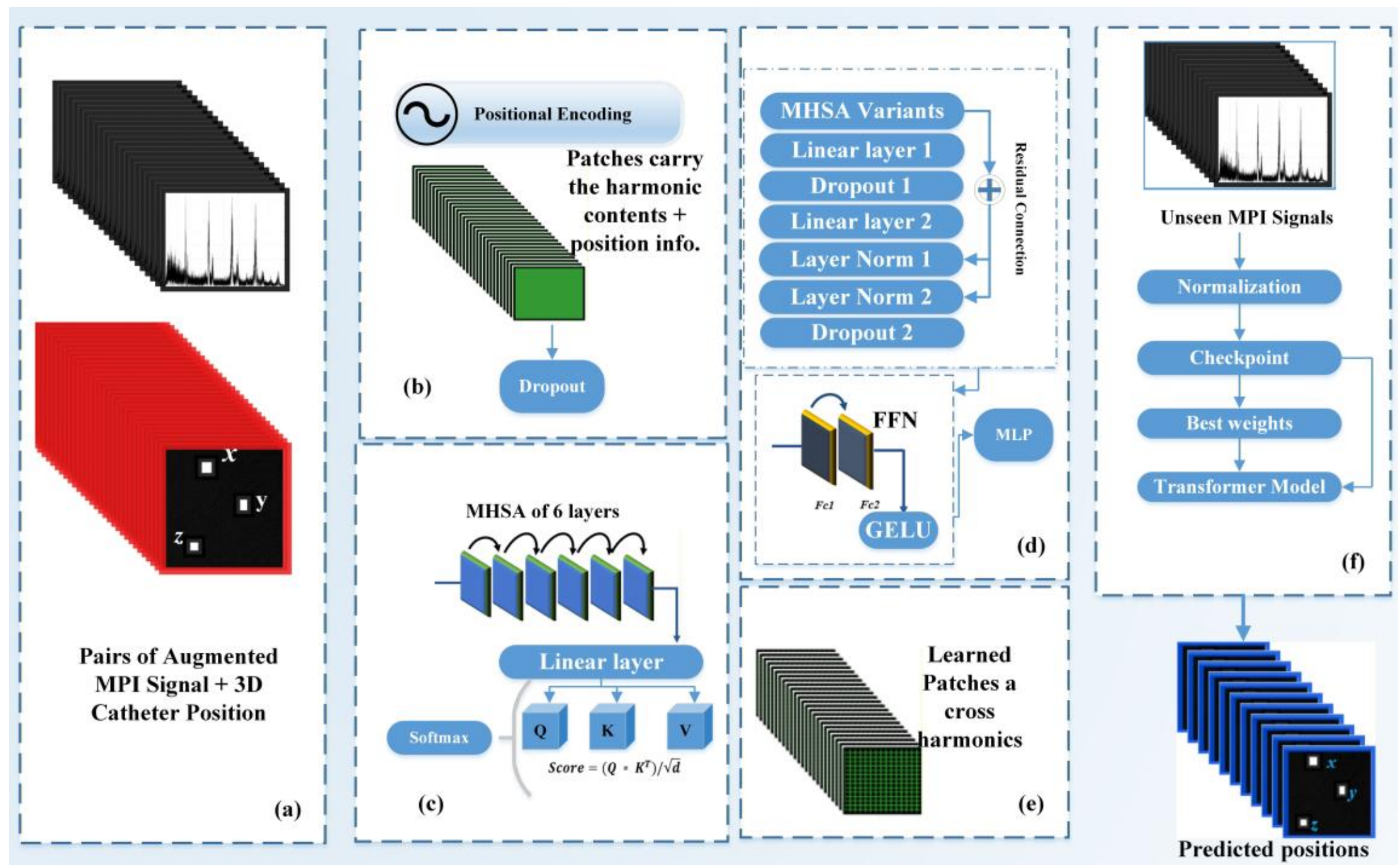

Figure 3: A schematic diagram of the proposed transformer model. (a) Pairs of MPI signals and catheter positions are used for training after preprocessing and augmentation through amplitude stretching, compression, and noise addition across four dataset types. (b) Positional encoding embeds patch-wise harmonic content along with positional information. (c) and (d) Multi-head self-attention computes scaled dot-product attention using query (Q), key (K), and value (V) projections with softmax normalization over the sequence length. The MPI signal is transformed into a global representation where informative harmonics for position estimation are emphasized. (e) Learned patch representations. (f) During inference, the trained model predicts catheter positions from unseen MPI signals.

### *2.2 Model Training*

The transformer network was trained using a configuration designed for stable and accurate catheter tip localization. The AdamW optimizer is used with a learning rate of $5 \times 10^{-5}$ and a weight decay of $5 \times 10^{-2}$. A 10-epoch warm-up phase is applied, followed by a cosine annealing schedule with warm restarts to improve convergence and avoid local minima.Training is performed for 200 epochs with a batch size of 128 and mixed-precision acceleration. Gradient clipping is used to prevent unstable updates. The model is trained using the weighted loss function defined in equation (14), where a higher weight is assigned to the Z-axis ($w_z$) to compensate for its lower spatial resolution. Performance is monitored using average loss, axis-wise MAE, and 3D localization error. The model achieving the lowest validation loss was automatically saved, and early stopping was used to prevent overfitting. All experiments were implemented in PyTorch on an Ubuntu 20.04.6 LTS system using four GPUs (Model: NVIDIA A100-SXM4-40GB with memory 40 GB HBM2e per GPU) .This configuration ensures efficient training and enables the model to learn a reliable mapping from frequency-domain MPI signals to accurate 3D catheter tip positions.

In contrast, conventional MPI reconstruction methods typically involve iterative algorithms such as system matrix inversion or regularized solvers, which introduce significantly higher computational latency, often on the order of several seconds per frame depending on the reconstruction settings. By directly mapping MPI signals to spatial coordinates, the proposed approach eliminates the reconstruction step, resulting in a substantial reduction in processing time and enabling real-time tracking performance.

### *2.3 Dataset and the experimental design*

Two dataset categories were used to develop and evaluate the proposed transformer-based catheter localization framework: a synthetic dataset generated through numerical simulation and experimental in vitro datasets reported in(Griese *et al*, 2020). The synthetic MPI dataset was produced using a numerical framework that models the magnetization dynamics of superparamagnetic nanoparticles under a time-varying magnetic field. The simulation employed a 1D drive field at 75 kHz with a peak amplitude of 10 mT, a spatial gradient of 10 T/m in all directions, and a 3 mT bias field for focusing. Nanoparticles with a mean diameter of 25 nm were simulated using the Langevin model, incorporating both N´eel and Brownian relaxation mechanisms.

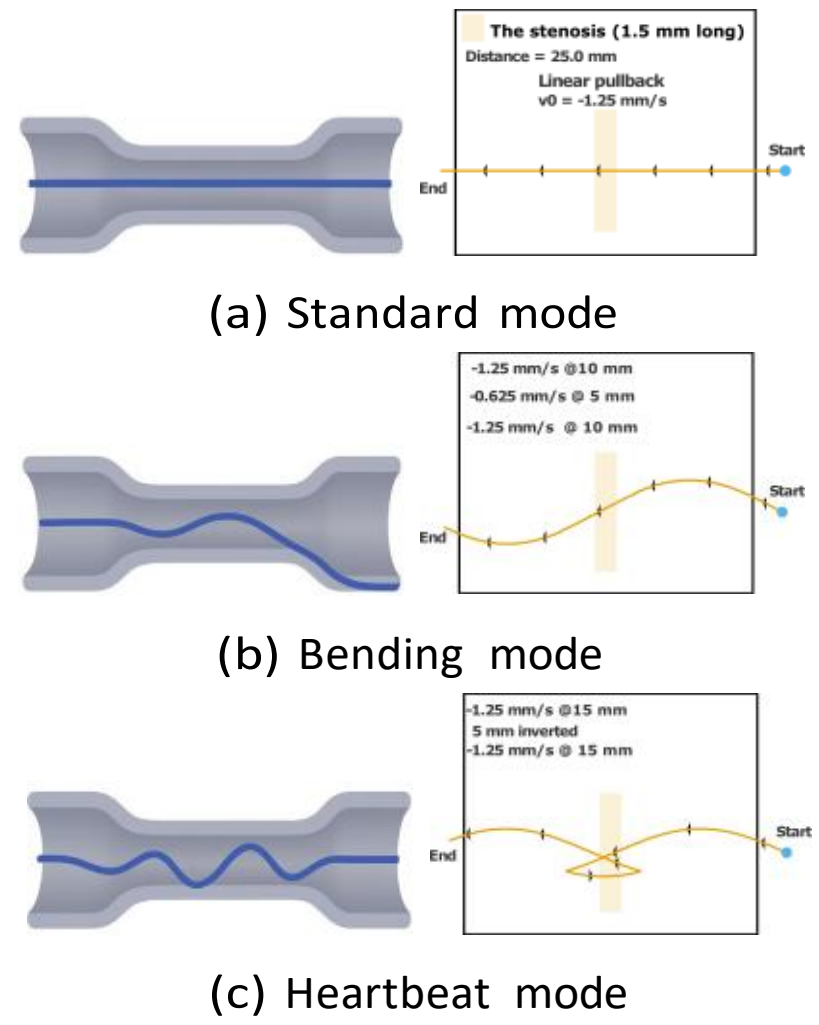

(a) Standard mode

(b) Bending mode

(c) Heartbeat mode

Figure 4: Experimental dataset catheter trajectories within phantom shown side by side: (a) standard, (b) bending, and (c) heartbeat modes as described in (Griese *et al* , 2020). Gray shaded region: 3D-printed phantom with stenosis; blue curves: catheter tip paths. Parameters: $v_0 = -1.25$ mm/s, pullback distance = 25 mm, repeats = 3, MPI delay = 1 s.

A smooth nonlinear catheter trajectory was generated using spline interpolation. For each of 8,000 spatial positions, an MPI signal with 80,000 time points was computed, including both real and imaginary components. To emulate realistic conditions, colored Gaussian noise (correlation 0.9, amplitude 0.05) was added to represent electronic and thermal noise. The experimental datasets used in this study were derived from the in-vitro setup described in (Griese *et al*, 2020), where a pre-clinical MPI scanner was combined with IVOCT system to track a magnetically coated catheter tip. A straight 3D-printed vessel phantom with a 1.5 mm stenosis was placed inside the MPI field of view, and a custom IVOCT pullback adapter was used to generate three distinct motion profiles. These profiles form the three datasets used in this work: standard (SP), bending (BA), and heartbeat (HBA) modes.

In the standard mode, the catheter was pulled back linearly at a constant velocity over a fixed distance, providing an artifact-free dataset. This mode serves as a baseline for calibration and evaluation under stable and uniform motion conditions.

In the bending mode, catheter deformation was simulated by varying the pullback velocity. Specifically, the catheter moved at a constant speed for the first 10 mm, slowed down over the next 5 mm, and then returned to its original velocity. This non-uniform motion introduces bending-related distortions similar to those observed due to friction or vessel curvature in practical scenarios.

In the heartbeat mode, physiological motion was emulated using a bidirectional trajectory. The catheter moved backward by 15 mm, forward by 5 mm, and then backward again by 15 mm. This repeated forward–backward pattern mimics cardiac-induced displacement, resulting in periodic motion artifacts commonly encountered in cardiovascular interventions.

These three datasets provide complementary motion conditions for evaluating the robustness and generalization capability of the proposed transformer-based framework.

### *2.4 Data extraction and preprocessing*

To enable real-time catheter tracking in MPI, a paired dataset of MPI signals and corresponding 3D catheter tip positions was constructed. A fully calibrated three-channel system matrix, stored in HDF5 format, served as the basis. Background frames were removed, and the matrix was reshaped to match the dimensionality of the measured MPI signals. Only frequency components consistent with the system matrix were retained, and frames with incompatible sizes were discarded to ensure data integrity.

Each signal frame was reconstructed into a 3D nanoparticle distribution using a regularized Kaczmarz solver. The catheter tip ground-truth position, denoted as $P_{\mathrm{tip}}$, was then extracted using a center-of-mass (CoM) approach (Griese *et al*, 2017). After thresholding the reconstructed volume to obtain a segmented region $V_{\text{seg}}$, the tip position was computed as:

$$P_{\text{tip}} = \frac{\sum_{x \in V_{\text{seg}}} I_{\text{seg}}(x, t)\, x}{\sum_{x \in V_{\text{seg}}} I_{\text{seg}}(x, t)}. \tag{16}$$

Table 1: Simulation parameters used for generating the synthetic MPI dataset.

| Parameter | Value | Unit |
|---|---|---|
| Number of samples | 8000 | – |
| Signal length | 80000 | points |
| Drive frequency | $75 \times 10^3$ | Hz |
| Drive field amplitude | $10 \times 10^{-3}$ | mT |
| Gradient field | [10.0, 10.0, 10.0] | T/m |
| Bias field | [0.003, 0.003, 0.003] | mT |
| Nanoparticle core size | $25 \times 10^{-9}$ | nm |
| Temperature | 300 | K |
| Coil sensitivity decay | $1 \times 10^{-3}$ | – |
| Time step | 1/(10 × drive frequency) | s |
| Signal noise level | 0.05 | a.u. |
| Magnetic constant ($\mu_0$) | $4\pi \times 10^{-7}$ | H/m |
| Boltzmann constant ($k_B$) | $1.38 \times 10^{-23}$ | J/K |
| Saturation magnetization ($M_S$) | 1.0 | a.u. |
| N´eel relaxation time | $1 \times 10^{-9}$ | s |
| Brownian relaxation time | $1 \times 10^{-8}$ | s |
| Colored noise correlation ($\alpha$) | 0.9 | – |

The segmented intensity $I_{\text{seg}}(x, t)$ is defined as:

$$I_{\text{seg}}(x, t) = \begin{cases} I(x, t), & \text{if } I(x, t) \geq \vartheta \cdot \max_x I(x, t), \\ 0, & \text{otherwise,} \end{cases} \tag{17}$$

where $\vartheta$ is a relative threshold. This procedure provides sub-voxel localization accuracy and ensures that the ground truth is consistently aligned with the MPI signal characteristics.

The resulting dataset consists of aligned signal–position pairs. For the simulated data, 8,000 pairs were generated (6,000 for training and 2,000 for inference). For the real datasets, 12,000 pairs were produced for both the standard and bending modes (9,000 for training/validation and 3,000 for testing each). The heartbeat mode contained 14,000 pairs (10,000 for training/validation and 4,000 for testing). This consistent and balanced data split ensures reliable evaluation across all motion profiles.

### *2.5 Preprocessing and Data Augmentation*

Before training the transformer model, a tailored preprocessing pipeline was applied to all MPI datasets to emphasize meaningful nanoparticle responses while suppressing noise and artifacts. Each MPI signal is recorded along the *x*-, *y*-, and *z*-receive coil channels and processed in the frequency domain using a harmonic-aware strategy.

The raw complex-valued signals,

$$S(t) = \{S_x(t), S_y(t), S_z(t)\}, \tag{18}$$

are first transformed using the Fast Fourier Transform (FFT), enabling harmonic decomposition. The fundamental excitation frequencies were approximately 24.5 kHz (x), 26.0 kHz (y), and 25.3 kHz (z), determined by the ratio between the drive amplitude and the scanner field-of-view.

To reduce spectral contamination, a band-pass filter was applied around the 2nd–8th harmonics of each fundamental frequency. The first harmonic was excluded because it is dominated by the superimposed modulation-field signal and provides limited catheter-related information (Gleich and Weizenecker, 2005; Bui *et al*, 2023; Janssen *et al*, 2021). The filtered signal from each axis is represented by its real and imaginary components and concatenated into a six-channel frequency-domain feature vector:

$$S(f) = \{(S_x, S_y, S_z)_r, \ (S_x, S_y, S_z)_i\}, \tag{19}$$

which preserves spatial and harmonic variations relevant to catheter localization.

All signals and their corresponding catheter positions were normalized using Min-Max scaling to the range [0, 1]. Signal normalization was applied after flattening and reshaping, while the 3D position labels were scaled independently for the training and validation splits to avoid data leakage. This preprocessing improves the signal-to-noise ratio (SNR) and standardizes the data across different acquisition modes.

To further enhance generalization, a lightweight data augmentation strategy was applied during training. For each frequency-domain signal $S \in \mathbb{R}^{C\times F}$, two randomized transformations were used:

**Amplitude scaling:** A scaling factor uniformly sampled from [0.9, 1.1] is applied element-wise to simulate coil sensitivity and gain variations.

**Noise perturbation:** Additive Gaussian noise corresponding to 5% of the mean absolute signal amplitude is injected to mimic realistic disturbances such as thermal noise or field drift. The resulting signal is clipped to prevent nonphysical values.

To maintain a consistent model input structure, a zero-valued padding channel was added where necessary, ensuring a standardized three-channel input ($x$, $y$, $z$) for all samples.

This domain-aware preprocessing and augmentation pipeline improves robustness to amplitude fluctuations, spectral noise, and minor experimental variations, enabling more stable and accurate catheter-tip localization across diverse MPI acquisition conditions.

**Algorithm 1: Harmonic-Based MPI Signal Preprocessing**

**Input:** $S \in \mathbb{C}^{3\times F}$ (complex MPI signals: $x, y, z$),
$P \in \mathbb{R}^3$ (true spatial positions),
$f_x, f_y, f_z$ (fundamental frequencies) [Hz],
$f_s$ (sampling frequency) [Hz],
$[h_1, h_2]$ (harmonic range)

**Output:** $\hat{S}$ (normalized harmonic-filtered signals), $\dot{P}$ (normalized positions),
$\{X_{\min}, X_{\max}, P_{\min}, P_{\max}\}$

**Procedure:**
$f_s \leftarrow 1/T_{\text{frame}}$
fundamentals $\leftarrow [f_x, f_y, f_z]$

**for each sample** $(S_i, P_i)$:
  $F_i \leftarrow \mathcal{F}\{S_i\}$(frequency-domain signal)
  **for each axis** $k \in \{x, y, z\}$:
    keep $|f - hf_k| \leq \Delta f, \ \forall h \in [h_1, h_2]$
    $F_i^{\text{filtered}}[k] \leftarrow F_i[k] \circ \text{mask}_k$
  $X_i \leftarrow [\text{Re}(F_i^{\text{filtered}}), \text{Im}(F_i^{\text{filtered}})]$
  store $(X_i, P_i)$

Fit Min–Max scalers:
$\hat{S} \leftarrow (X - X_{\min})/(X_{\max} - X_{\min})$
$\dot{P} \leftarrow (P - P_{\min})/(P_{\max} - P_{\min})$
Save $\hat{S}$, $\dot{P}$ and $\{X_{\min}, X_{\max}, P_{\min}, P_{\max}\}$

**Return:** Normalized, frequency-filtered MPI dataset
**Transformer-based localization:**
  Divide $\hat{S}$ into non-overlapping patches using 1D convolution
  Apply positional encoding to embedded patches
  Process tokens using multi-head self-attention encoder layers
  Extract global feature representation $Z_{\text{enc}}$
  Predict catheter coordinates using MLP regression head:
    $\hat{Y} = [\hat{x}, \hat{y}, \hat{z}]$

Optimize model using weighted MAE loss

**Return:** Predicted three-dimensional catheter tip position

**Parameters:**
$\mathcal{F}\{\cdot\}$ : Fourier transform operator
$\Delta f$: bandwidth around each harmonic
$\circ$: element-wise multiplication

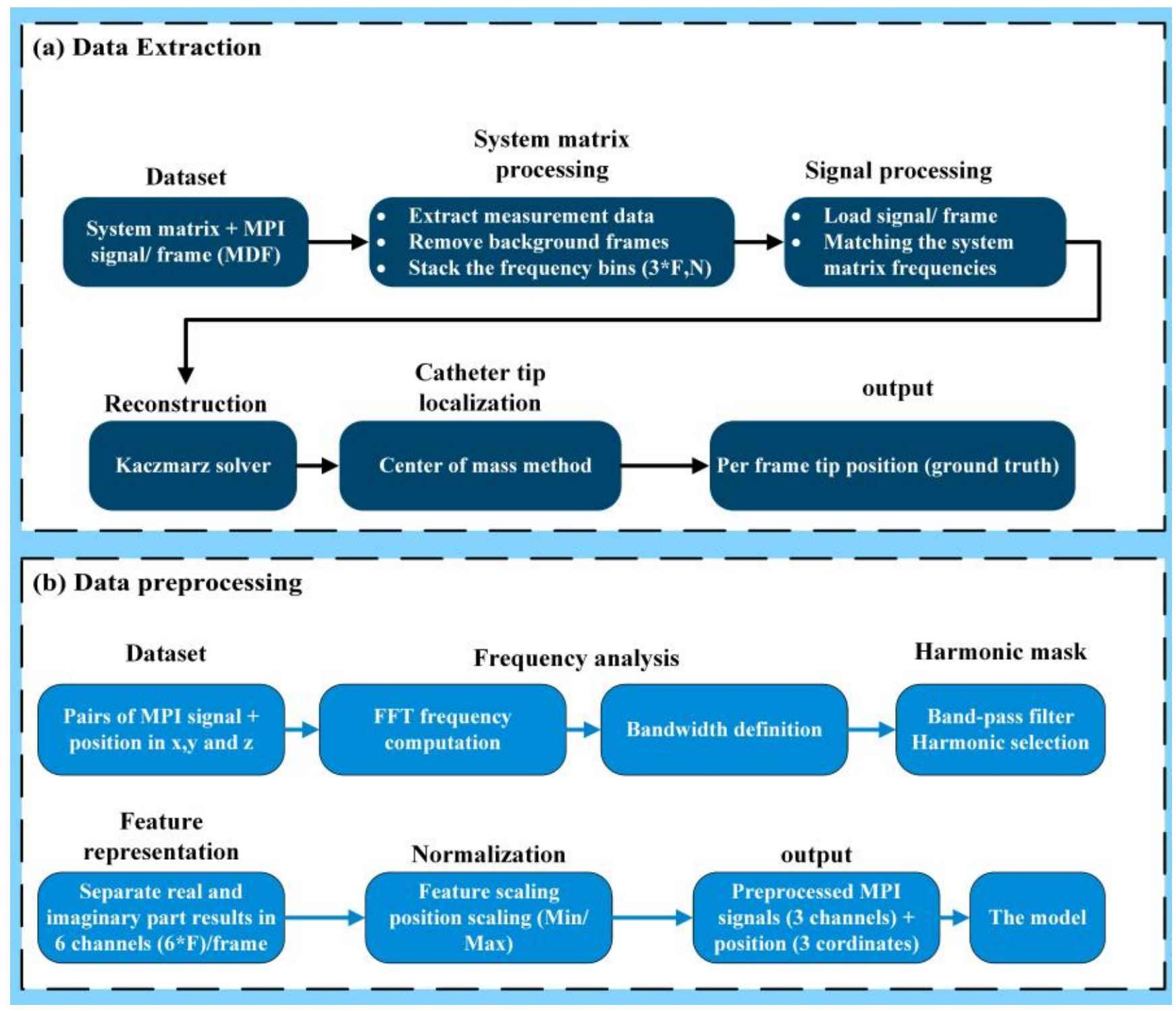

Figure 5: The data extraction (a) and preprocessing (b) steps. MPI data stored in MPI data format (MDF) format are converted into signal frames, from which catheter positions are reconstructed using the center-of-mass method to generate ground-truth labels. Each signal–position pair is then preprocessed based on the frame duration and excitation frequencies used during acquisition before being input to the model in Figure 3.

## 3 Results

### *3.1 Model performance*

Across the four datasets, the learning curves show distinct behaviors, as summarized in Figure 7. In the simulated dataset, the training loss decreases smoothly and the mean absolute error (MAE) improves steadily; however, mild fluctuations in the validation curves indicate that synthetic signals do not fully capture the complexity of real MPI data.

The standard mode achieves the most stable convergence, where both training and validation losses drop rapidly and align closely, with consistently low per-axis MAEs. The bending (BA) mode shows similarly strong learning behavior, with tightly matched training and validation losses and MAEs across the $X$, $Y$, and $Z$ axes decreasing to very low values. The L2 error also decreases rapidly and stabilizes, confirming reliable performance under bending motion.

In the heartbeat (HBA) mode, the losses decrease early during training, but the validation curve plateaus sooner, suggesting mild overfitting. Although both MAE and $L_2$ errors improve, they remain slightly higher than those of the BA mode due to the increased complexity introduced by heartbeat-induced motion.

### *3.2 Metrics evaluation*

Catheter-tip localization is evaluated using multiple complementary metrics. Axis-wise mean absolute error (MAE) along the $X$, $Y$, and $Z$ directions, the overall L2 error, and the root mean square error (RMSE) quantify spatial accuracy and variability. In addition, residual plots are used to analyze error distributions, while quantile–error curves summarize median, 90th-percentile, and worst-case behavior. Temporal tracking performance is further assessed using predicted versus ground-truth trajectories, frame-wise $L_2$ errors, and velocity errors to evaluate motion smoothness and dynamic accuracy.

As summarized in Table 3, the simulated dataset exhibits the largest errors, indicating that the model captures general motion trends but lacks fine precision under synthetic conditions. Performance improves significantly in the standard (SP) mode, reflecting more reliable predictions on real catheter signals. The bending (BA) mode achieves the best overall accuracy, demonstrating strong robustness to deformation dynamics. The heartbeat (HBA) mode performs between SP and simulated data, with slightly higher errors due to the added complexity of cardiac-like periodic motion.

Table 2: Imaging protocol for the experimental data.

| Parameter | Value | Unit / Notes |
|---|---|---|
| MPI frequencies | 2.5/102, 2.5/96, 2.5/99 | MHz |
| Drive field | 12 | mT |
| Gradient (z, x, y) | 2.0, 1.0, 1.0 | T/m |
| Frame rate | 46.43 | Hz |
| FOV | 24 × 24 × 12 | mm |
| Cropped FOV | 10 | mm |
| Calibration grid | 35 × 25 × 13 | positions |
| Calibration step | 1 × 1 × 1 | mm |
| Tracer volume | 1 | $\mu$L |
| Threshold ($\vartheta$) | 0.35 | – |
| Phantom ID | 2.5 | mm |
| Phantom length | 20 | mm |
| Stenosis | 1.5 × 1.5 | mm |
| A-scan pixels | 512 | samples |
| Catheter rotation | 6.25 | Hz |
| Catheter OD | 0.9 | mm |
| Pullback velocity ($v_0$) | −1.25 | mm/s |
| Pullback distance | 25 | mm |
| **Motion profiles:** | | |
| SP | Linear | $v_0$ |
| BA | 10/5/10 mm | $v_0/0.5v_0/v_0$ |
| HBA | 15/5/15 mm | $v_0/-v_0/v_0$ |
| Repeats | 3 | times |

Table 3: Performance metrics comparing predicted and ground-truth catheter positions for all datasets (mean ± STD).

| Dataset | MAE(X) (mm) | MAE(Y) (mm) | MAE(Z) (mm) | L2 Error (mm) |
|---|---|---|---|---|
| Simulated | 0.282 ± 0.042 | 0.282 ± 0.059 | 0.337 ± 0.031 | 0.595 ± 0.088 |
| Standard | 0.133 ± 0.103 | 0.130 ± 0.103 | 0.089 ± 0.059 | 0.229 ± 0.167 |
| Bending | 0.039 ± 0.046 | 0.054 ± 0.049 | 0.060 ± 0.044 | 0.103 ± 0.092 |
| Heartbeat | 0.107 ± 0.076 | 0.081 ± 0.065 | 0.099 ± 0.077 | 0.185 ± 0.13 |

These trends are further confirmed by the quantile–error curves (see supplementary materials, Section 5). The simulated dataset shows the widest error spread (median $\approx$ 1.12 mm, maximum $\approx$ 1.65 mm), whereas the SP mode exhibits a tighter distribution (median $\approx$ 1.02 mm, maximum $\approx$ 1.34 mm). The BA mode again demonstrates the best performance, with a narrow error distribution (median $\approx$ 0.33 mm, maximum $\approx$ 0.39 mm). In contrast, the HBA mode shows a moderate median error ($\approx$ 0.79 mm) but a larger maximum deviation ($\approx$ 2.81 mm), reflecting its more irregular temporal dynamics.

While quantile analysis provides a global view of error magnitude, residual analysis offers insight into directional error behavior. As illustrated in Figure 8, the simulated dataset exhibits the widest and most scattered residuals, indicating higher variability and reduced precision. The SP mode shows a tighter, zero-centered distribution, suggesting stable and unbiased predictions. The BA mode presents the most compact clustering, confirming highly accurate tracking under bending motion. The HBA mode displays a broader residual spread due to its dynamic and irregular motion patterns. These observations indicate that the model remains largely unbiased while adapting to varying motion complexities.

Temporal tracking performance is illustrated in Figure 9, which compares predicted and ground-truth trajectories over time. The shaded region represents frame-wise errors: narrow regions indicate close agreement, whereas wider regions reveal drift. This visualization highlights both the accuracy and temporal consistency of the proposed framework.

The fourth plot in each group in Figure 9 (red) shows the frame-wise L2 error, representing the Euclidean distance between predicted and ground-truth positions in mm. The solid curve denotes instantaneous error, while the dashed line and shaded band indicate the mean $\pm$ one standard deviation. These plots confirm stable predictions with occasional spikes, particularly in more complex motion scenarios such as the HBA dataset.

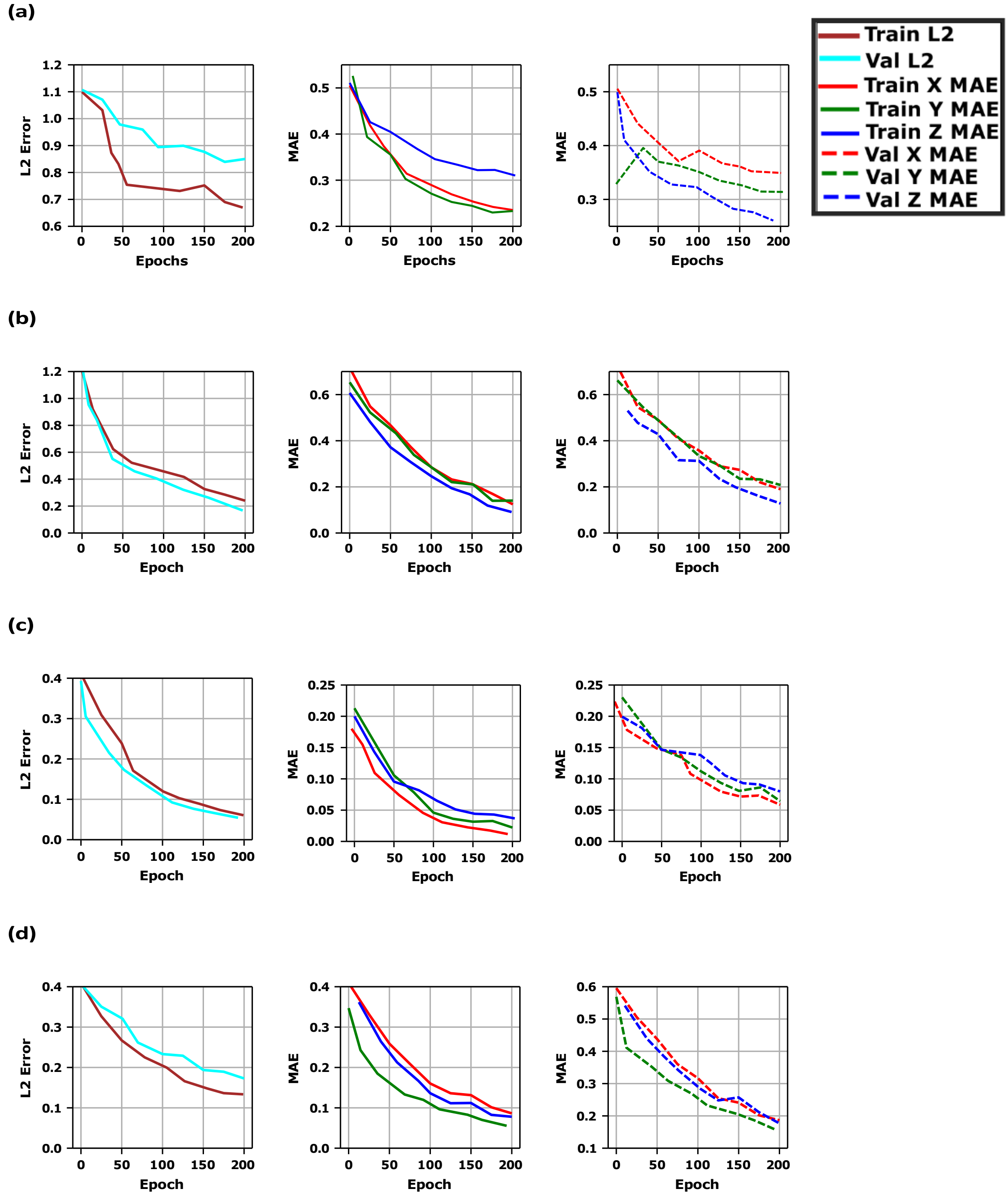

Figure 6: Model performance across 200 training epochs: (a) simulated, (b) SP, (c) BA, and (d) HBA datasets. Each row shows L2 error, training MAE (X,Y,Z), and validation MAE (X,Y,Z) evolution respectively.

### 3.3 *Evaluation of localization accuracy*

To evaluate the proposed framework and confirm the results from section 3.2, we analyzed localization accuracy across all datasets, focusing on both precision and robustness under different motion scenarios. Cumulative thresholds ( $\leq$0.25 mm, $\leq$0.5 mm, $\leq$1 mm, $\leq$2 mm) quantify the proportion of frames within clinically relevant bounds. The findings are summarized in Tables 4 and 5.

Table 4 summarizes the percentile analysis of localization errors across all frames for each dataset. The simulated dataset shows a slightly wider spread, with median (P50) at 1.1 mm increasing to 1.6 mm at the 99th percentile, indicating occasional higher deviations. The standard dataset is highly stable, rising only from 1.0 mm to 1.04 mm across percentiles. The bending dataset exhibits the highest consistency, with values nearly constant at 0.33–0.34 mm, reflecting uniform sub-millimeter accuracy. The heartbeat dataset shows intermediate behavior, with median errors around 0.79 mm and a slight rise to 0.81 mm at the 99th percentile, maintaining tight error bounds even under dynamic motion.

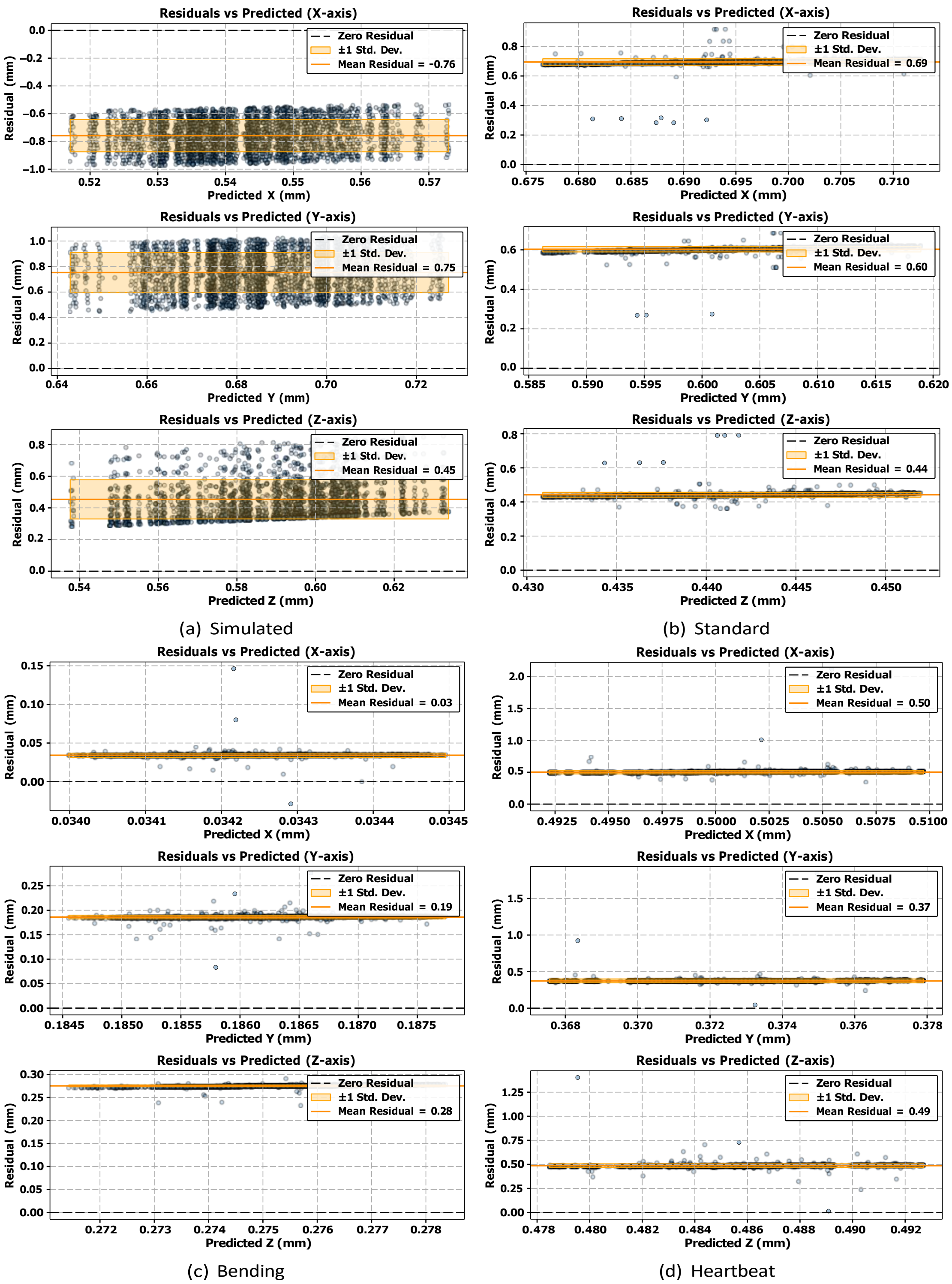

(a) Simulated (b) Standard

(c) Bending (d) Heartbeat

Figure 7: Residual errors of the predicted catheter positions across all axes ($x$, $y$, and $z$): (a) simulated data, (b) standard protocol (SP), (c) bending (BA), and (d) heartbeat (HBA).

Table 5 presents the cumulative threshold evaluation, showing how many frames fall within clinically relevant error bounds. In the simulated dataset, no frames were below 0.25 mm or 0.5 mm, while 29% were below 1 mm, and all frames (100%) were within 2 mm. The standard dataset shows similar behavior at finer thresholds (0–0.5 mm), with only 3.3% of frames below 1 mm, but again full coverage within 2 mm.

The bending dataset exhibits the best performance, with all frames achieving sub-millimeter accuracy (100% $\leq$ 0.5 mm), indicating highly stable localization even under bending motion. The heartbeat dataset also performs strongly, with 99.9% of frames below 1 mm and 99.97% below 2 mm, demonstrating reliable accuracy under dynamic physiological motion.

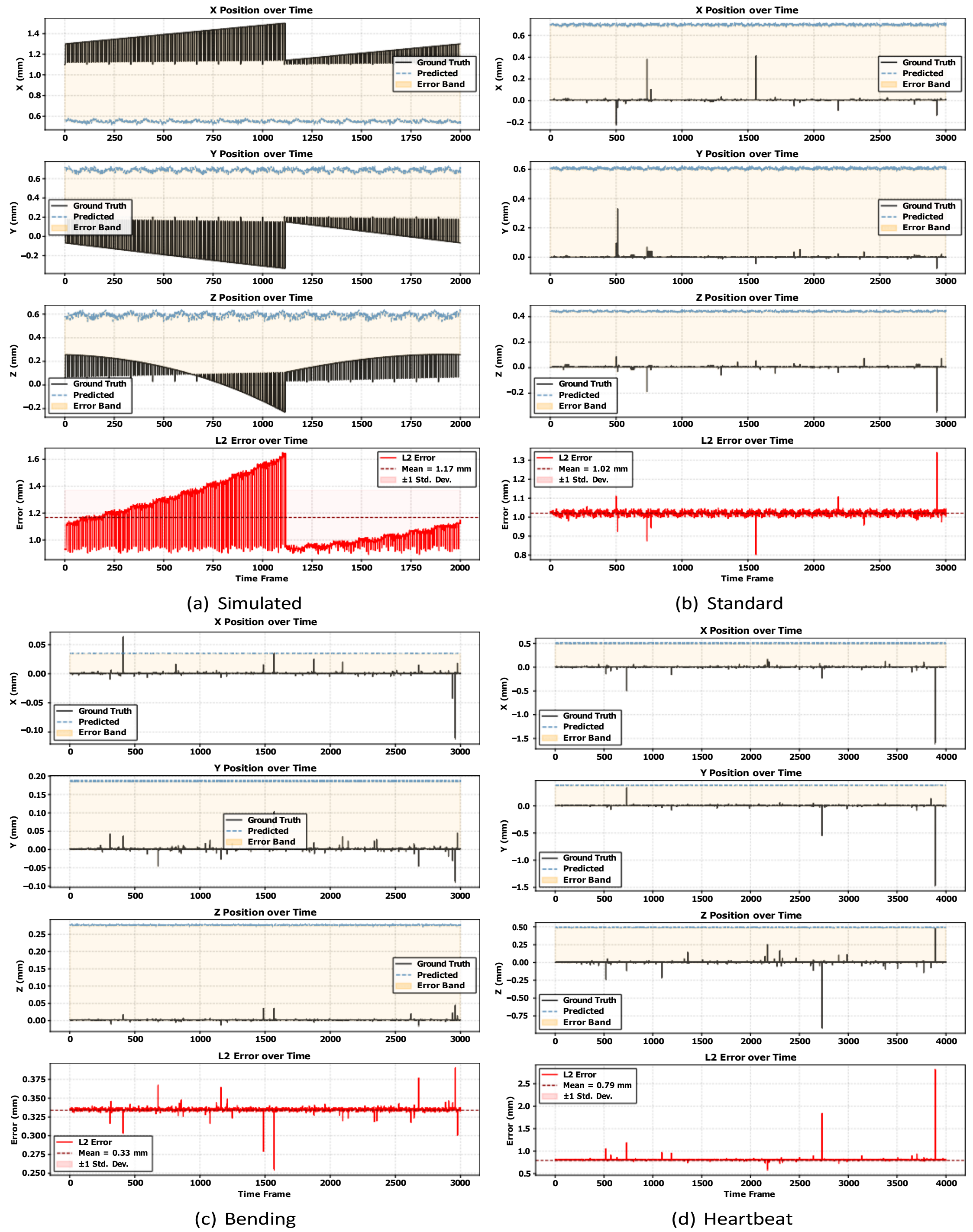

(a) Simulated (b) Standard

(c) Bending (d) Heartbeat

Figure 8: Temporal resolution evaluation showing predicted versus ground-truth catheter positions along the *x*, *y*, and *z* axes over time. The L2 error over time (mean ± standard deviation) is shown in red. Each catheter position corresponds to one frame: (a) simulated data, (b) standard protocol (SP), (c) bending (BA), and (d) heartbeat (HBA).

### 3.4 *Latency and footprint*

We assessed the efficiency of the proposed framework using several runtime metrics, including latency, throughput, FLOPs, and inference time on unseen MPI signals. These metrics jointly characterize responsiveness, computational load, and real-time suitability.

In contrast, the optimized inference model (Table 7) achieved substantial gains. Batch latency was reduced to ∼69 ms and frame latency to ∼0.55 ms, resulting in a throughput of ∼1800 frames/s. Total inference time remained low across datasets, ranging from 4.785 s

Table 4: Percentile analysis of localization error across the testing datasets.

| Dataset | P50 | P55 | P90 | P95 | P99 |
|---|---|---|---|---|---|
| Simulated | 1.119 | 1.316 | 1.483 | 1.553 | 1.612 |
| Standard | 1.021 | 1.029 | 1.035 | 1.038 | 1.044 |
| Bending | 0.330 | 0.340 | 0.340 | 0.340 | 0.340 |
| Heartbeat | 0.791 | 0.795 | 0.799 | 0.801 | 0.810 |

Table 5: Cumulative threshold ($T$) analysis of localization error across the frames of testing datasets.

| Dataset | $T$ (mm) | $\leq T$ | Total | (%) |
|---|---|---|---|---|
| Simulated | 0.25 | 0 | 2000 | 0 |
| | 0.5 | 0 | 2000 | 0 |
| | 1 | 582 | 2000 | 29.1 |
| | 2 | 2000 | 2000 | 100 |
| Standard | 0.25 | 0 | 3000 | 0 |
| | 0.5 | 0 | 3000 | 0 |
| | 1 | 100 | 3000 | 3.33 |
| | 2 | 3000 | 3000 | 100 |
| Bending | 0.25 | 0 | 3000 | 0 |
| | 0.5 | 3000 | 3000 | 100 |
| | 1 | 3000 | 3000 | 100 |
| | 2 | 3000 | 3000 | 100 |
| Heartbeat | 0.25 | 0 | 4000 | 0 |
| | 0.5 | 0 | 4000 | 0 |
| | 1 | 3996 | 4000 | 99.9 |
| | 2 | 3999 | 4000 | 99.97 |

Table 6: Model footprint metrics using training dataset.

| Metric | Value |
|---|---|
| Batch latency (s) | 1.507 |
| Frame latency (ms) | 11.78 |
| Throughput (frame/s) | 84.91 |
| FLOPs / Sample (Mega) | 8.81 |
| FLOPs / Batch (Mega) | 1127.41 |
| Total FLOPs (Giga) | 56.370 |

Table 7: Footprint metrics for inference model on testing dataset.

| Dataset | Batch (ms) | Frame (ms) | Throughput (frame/s) | Inference (s) | RMSE (mm) | Velocity error* (mm/frame) |
|---|---|---|---|---|---|---|
| Simulated | 69.722 | 0.558 | 1792 | 4.785 | 0.683 | 0.0838 |
| Standard | 69.087 | 0.533 | 1809 | 6.326 | 0.589 | 0.0090 |
| Bending | 69.178 | 0.553 | 1806 | 6.362 | 0.192 | 0.0023 |
| Heartbeat | 69.345 | 0.555 | 1802 | 8.247 | 0.458 | 0.0090 |

*Velocity error is explained in Supplementary Materials, Section 6.

(Simulated) to 8.247 s (HBA data). Notably, this processing speed is significantly faster than the MPI hardware limit (Table 6), which provides a new signal frame approximately every 43 ms (23.2 Hz), confirming the model's suitability for real-time interventional use.

### *3.5 Ablation study*

To assess the contribution of key components in our framework, we conducted ablation studies on the Heartbeat dataset. Each setting was trained for 200 epochs, and the best-performing weights were evaluated using the metrics reported in Table 8. The results of the three ablation categories are summarized below.

**(1) Harmonic Range Selection:** We evaluated how harmonic bands affect localization by isolating low (2nd–4th), high (5th–8th), and very high (9th–20th) orders using FFT filtering. Lower harmonics provide strong coarse features, while higher ones encode finer details. Results are reported in Table 8.

**(2) Frame Duration:** Two temporal windows were tested: a short 0.20 ms frame (high temporal resolution) and a long 0.05 s frame (noise smoothing), both using the 2nd–8th harmonics. Their effect on reconstruction accuracy is summarized in Table 8.

**(3) Model Architecture:** We compared architectural variants, testing models with 4 vs. 16 attention heads and shallow (8-layer) vs. deeper (12-layer) transformers. These experiments assess the trade-off between accuracy, efficiency, and real-time feasibility.

**(4) Effect of Z-Axis Weighting in Loss Function:** We also examined the impact of a design choice related to the loss function. Recall that the z-axis typically suffers from lower spatial resolution compared to the x and y axes. To compensate for this limitation, our baseline model applies a higher weight to the z-axis loss term during training. In this ablation, we removed that special weighting and retrained the model under identical conditions as shown in Table 8.

Table 8: Ablation study on Heartbeat mode dataset. The MAE and L2 metrics are reported from the training data, while the RMSE, velocity error, inference time, and throughput are computed using the inference model on the testing data.

| Ablation Study | MAE (mm) X | MAE (mm) Y | MAE (mm) Z | L2 Error (mm) | RMSE (mm) | Velocity Error (mm/frame) | Inference Time (s) | Throughput (frame/s) |
|---|---|---|---|---|---|---|---|---|
| Higher order harmonics | 0.090±0.052 | 0.084±0.061 | 0.094±0.066 | 0.173±0.117 | 0.557 | 0.0090 | 11.017 | 939 |
| Lower order harmonics | 0.112±0.093 | 0.088±0.072 | 0.109±0.087 | 0.199±0.158 | 0.519 | 0.0080 | 11.317 | 990 |
| 9th–20th harmonics | 0.114±0.093 | 0.075±0.052 | 0.096±0.063 | 0.185±0.136 | 0.522 | 0.0093 | 10.781 | 1491 |
| Long frame duration | 0.098±0.070 | 0.073±0.050 | 0.098±0.078 | 0.173±0.128 | 0.470 | 0.0090 | 10.164 | 1012 |
| Short frame duration | 0.104±0.078 | 0.080±0.059 | 0.086±0.052 | 0.173±0.123 | 0.470 | 0.0094 | 10.731 | 918 |
| 4 attention heads | 0.110±0.078 | 0.082±0.062 | 0.103±0.072 | 0.190±0.136 | 0.529 | 0.0094 | 10.154 | 1237 |
| 16 attention heads | 0.099±0.064 | 0.085±0.067 | 0.102±0.073 | 0.184±0.131 | 0.573 | 0.0090 | 12.661 | 704 |
| 8 transformer layers | 0.098±0.068 | 0.087±0.065 | 0.099±0.067 | 0.182±0.128 | 0.472 | 0.0087 | 11.818 | 742 |
| 12 transformer layers | 0.106±0.072 | 0.084±0.056 | 0.090±0.056 | 0.181±0.120 | 0.455 | 0.0080 | 10.601 | 993 |
| Unweighted loss | 0.143±0.093 | 0.127±0.084 | 0.133±0.079 | 0.263±0.158 | 0.869 | 0.0002 | 9.746 | 1231 |
| Proposed Model | 0.107±0.076 | 0.081±0.065 | 0.099±0.077 | 0.185±0.139 | 0.458 | 0.0090 | 8.247 | 1802 |

Table 9: Comparison of selected real-time MPI studies showing accuracy, performance, and key contributions.

| **Study** | **Accuracy/Error Metrics** | **Performance/Real-time Capability** | **Key Contribution** |
|---|---|---|---|
| MPI/MRI-Guided Angioplasty Salamon et al. (2016) | Voxel resolution: 2×2×1 mm³ | Online: 2 vol/s, Full: 46 vol/s | First 4D real-time in-vitro MPI-guided angioplasty. |
| MPI Catheter Steering Rahmer *et al* (2017) | Sub-mm visual accuracy | Online: 5 Hz, Full: 46 Hz | First multi-color MPI with magnetic catheter steering. |
| Sub-mm Marker Localization Griese *et al* (2017) | Error: 0.27 mm (x,y), 0.12 mm (z) | 46 Hz, 1.3 ms/frame | Center-of-mass-based marker localization in MPI. |
| MPI-Guided PTA Herz *et al* (2018) | Visual confirmation | 4 fps, 100 ms latency | First real-time MPI-guided angioplasty in phantom model. |
| MPI-guided IVOCT Tracking Griese *et al* (2020) | MAE: 0.2 ± 0.22 mm | 23.2 Hz real-time | Motion artifact compensation in MPI-guided IVOCT tracking. |
| MPI Balloon Catheter Ahlborg et al. (2022) | – | 2 fps | First dedicated balloon catheter for MPI. |
| AMPIM Shen *et al* (2023) | Reconstruction: 98–99% faster | 0.3–2.2 ms/frame | GPU-accelerated real-time MPI reconstruction framework. |
| DL for MPI Resolution Wen *et al* (2024) | RMSE: 0.022 | Post-processing | CNN with prior knowledge embedding for MPI super-resolution. |
| Bare-Metal Stent Tracking Wegner *et al* (2024) | MAE: 1.46–1.49 mm | – | Tracking unmodified bare-metal stents in MPI. |
| CAD-Net Shi *et al* (2025) | PSNR / SSIM (qualitative) | 78.46 fps (sim), 67.85 fps (real) | Content-aware distillation for real-time MPI imaging. |
| MIFT-Net Sun *et al* (2025) | SNR / PSNR (qualitative) | 28.52 ms/signal | Transformer-based signal denoising for MPI. |
| **Proposed Method** | MAE: 0.039 ± 0.092 mm | 0.55 ms/frame (1800 fps) | Real-time catheter tracking using transformer-based signal modeling. |

Table 9 summarizes representative real-time MPI studies and highlights the progressive development of the field in terms of accuracy, computational speed, and application scope. Overall, early works primarily focused on demonstrating the feasibility of MPI in interventional settings, achieving either acceptable real-time performance or sufficient spatial accuracy, but rarely both simultaneously. For instance, MPI-guided angioplasty and four-dimensional imaging approaches demonstrated real-time capabilities at the system level Salamon et al. (2016), while MPI-based catheter steering extended this toward multi-color tracking and magnetic actuation Rahmer *et al* (2017). Similarly, marker-based localization methods achieved sub-millimeter precision (e.g., 0.27 mm in the lateral plane), but under highly controlled conditions with simplified experimental setups Griese *et al* (2017). Early clinical translation efforts such as MPI-guided percutaneous transluminal angioplasty further confirmed feasibility in phantom environments, albeit with limited temporal resolution and latency constraints Herz *et al* (2018).

More recent studies demonstrate a clear transition toward computational acceleration and data-driven modeling. Motion-compensated and tracking-based approaches, such as MPI-guided IVOCT systems, improved localization accuracy while maintaining real-time frame rates around 20–25 Hz Griese *et al* (2020). Hardware-specific innovations, including dedicated balloon catheter designs, further enhanced system integration for interventional compatibility, though real-time performance remained moderate Ahlborg et al. (2022). On the computational side, GPU-accelerated frameworks such as an adaptive multi-frame parallel iterative method (AMPIM) significantly reduced reconstruction time to the sub-millisecond range, achieving reconstruction speedups of up to two orders of magnitude Shen *et al* (2023). Deep learning-based methods further advanced image quality and speed, with CNN-based super-resolution models improving RMSE performance Wen *et al* (2024), while recent content-aware distillation networks achieved real-time imaging rates exceeding 60 fps in experimental settings Shi *et al* (2025). Transformer-based signal enhancement approaches have also been introduced, improving signal to noise ratio (SNR) and peak signal to noise ratio (PSNR) through learned denoising strategies, although often evaluated indirectly and not always tied to direct localization accuracy Sun *et al* (2025).

In contrast, the proposed method achieves a significantly improved balance between accuracy and computational efficiency, reaching a mean absolute error of 0.039 $\pm$ 0.092 mm while maintaining ultra-low latency of 0.55 ms per frame (approximately 1800 fps). This indicates that signal-level transformer modeling can effectively bridge the gap between high-speed reconstruction methods and high-precision localization frameworks, addressing a key limitation consistently observed across prior MPI studies.

## 4 Discussion

This study introduces a signal-only deep learning framework for real-time catheter tip localization in MPI, demonstrating that accurate spatial information can be recovered directly from frequency-domain voltage signals without explicit image reconstruction. By combining harmonic-aware preprocessing with a transformer architecture, the proposed method achieves high localization accuracy while maintaining low computational cost, making it suitable for real-time interventional applications.

Across datasets, the model exhibits strong generalization and robustness to different motion patterns. In particular, the bending (BA) mode consistently outperforms other datasets. This behavior can be attributed to the increased spectral richness introduced by nonlinear motion. The BA trajectory modulates the field-free point (FFP) velocity and gradient fields, generating broader harmonic excitation and stronger cross-axis coupling. As a result, the input signals contain more informative, position-dependent spectral features, enabling the transformer's self-attention mechanism to better resolve spatial variations. This effect can be interpreted as increased "spectral diversity," which enhances model learnability and robustness to noise (Panagiotopoulos *et al*, 2015; Bui *et al*, 2023; Rahmer *et al*, 2009; Top *et al*, 2019). The larger maximum deviation observed in the heartbeat (HBA) mode can be attributed to the inherently irregular motion pattern, which includes rapid direction changes and non-uniform velocities. Unlike the smoother trajectories in the standard and bending modes, this dynamic behavior introduces more complex and less predictable signal variations, making accurate localization more challenging. Although the proposed model maintains good overall accuracy, these abrupt temporal fluctuations can occasionally lead to higher peak errors.

The harmonic-aware preprocessing strategy plays a critical role in achieving this performance. By focusing on the 2nd–8th harmonics, the framework captures the most stable and informative MPI signal components while suppressing noise-dominated higher-order harmonics. This selection balances spatial resolution and noise robustness, consistent with prior studies on harmonic signal encoding in MPI (Rahmer *et al*, 2009). The ablation results further confirm that combining low- and mid-order harmonics provides superior performance compared to using either range alone.

The 2nd–8th harmonics were selected to balance spatial information and robustness. Lower-order harmonics provide stronger signals with higher signal-to-noise ratio (SNR), while mid-order harmonics add useful spatial detail. Although higher-order harmonics may contain finer information, their amplitudes are much weaker and more sensitive to noise, which limits their practical benefit. We also tested higher-order harmonics and observed no improvement in localization accuracy, with some degradation due to noise. These results support the chosen range as an effective trade-off. Additional details are provided in the supplementary materials section 3.

Compared to conventional MPI reconstruction methods (as in Table 9), which typically rely on iterative solvers and can require substantial computational time per frame, the proposed approach eliminates the reconstruction step by directly mapping signals to spatial coordinates. This results in significantly reduced latency and enables real-time tracking performance, as demonstrated by the high throughput reported in Table 8.

Despite these promising results, several limitations should be considered. First, the model is data-driven and may be sensitive to domain shifts. The higher errors observed in simulated datasets highlight the gap between idealized signal generation and real MPI measurements, where factors such as coil sensitivity variations, electronic noise, and phase inconsistencies introduce additional complexity. Second, the experimental validation is limited to a single MPI system and phantom configuration, which may restrict generalizability across different hardware setups and anatomical scenarios. Third, the regression-based formulation does not explicitly enforce physical constraints, which may occasionally lead to unrealistic predictions under extreme conditions.

Future work will focus on addressing these limitations by extending validation to multiple MPI systems and phantoms, incorporating more realistic simulation models, and performing in vivo experiments under physiological motion. In addition, integrating uncertainty-aware prediction mechanisms and physics-informed constraints could further improve robustness and reliability. Finally, optimizing the model for hardware-efficient deployment will be essential for translation into clinical environments.

## 5 Conclusion

This work presents a lightweight transformer-based framework for real-time catheter tip localization in MPI. By operating directly on frequency-domain signals and leveraging harmonic-aware preprocessing, the proposed method achieves sub-millimeter accuracy with inference speeds exceeding 1,800 frames/s.

The results demonstrate that accurate spatial localization can be achieved without image reconstruction, significantly reducing latency while maintaining high precision. The framework shows strong robustness across different motion patterns, with improved performance linked to the spectral richness of the input signals.

These findings highlight the potential of signal-domain learning for fast and reliable MPI-guided interventions. Future developments in in vivo validation, uncertainty modeling, and hardware optimization will further advance the applicability of this approach in real-time clinical settings.

**Acknowledgments**

This work was supported in part by the National Key R&D Program of China (Grant No. 2025YFC2426200), the National Natural Science Foundation of China (Grant No. 82227802), and the Taishan Industrial Experts Program.

**Data availability**

The experimental datasets used in this study are publicly available at https://zenodo.org/records/3554935. These datasets were originally introduced in Griese *et al* (2020).

**Supplementary data**

Supplementary material associated with this manuscript is attached and available for review.

**References**

Ahlborg M *et al* First dedicated balloon catheter for magnetic particle imaging *IEEE Trans. Med. Imaging* **41**(11) 3301–3308 doi:10.1109/TMI.2022.3183948

Bui T Q *et al* 2023 Harmonic dependence of thermal magnetic particle imaging *Sci. Rep.* **13**(1) doi:10.1038/s41598-023-42620-1

Gapyak V *et al* 2025 An $\ell_1$-plug-and-play approach for MPI using a zero-shot denoiser *Phys. Med. Biol.* **70**(2) doi:10.1088/1361-6560/ada5a1

Gleich B 2005 Tomographic imaging using the nonlinear response of magnetic particles *Nature* **435**(7046) 1214–1217 doi:10.1038/nature03808

Griese F *et al* 2017 Submillimeter-accurate marker localization within low gradient magnetic particle imaging tomograms *Int. J. Magn. Part. Imaging* **3**(1) doi:10.18416/ijmpi.2017.1703011

Griese F *et al* 2020 In-vitro MPI-guided IVOCT catheter tracking in real time for motion artifact compensation *PLoS One* **15**(3) doi:10.1371/journal.pone.0230821

Guo L S *et al* 2025 Exploring the diagnostic potential: magnetic particle imaging for brain diseases *Mil. Med. Res.* **12**(1) doi:10.1186/s40779-025-00603-5

Haegele J *et al* 2012 Magnetic particle imaging: Visualization of instruments for cardiovascular intervention *Radiology* **265** doi:10.1148/radiol.12120424

Han X *et al* 2020 The applications of magnetic particle imaging: From cell to body *Diagnostics* **10**(10) 800 doi:10.3390/diagnostics10100800

Hartung V *et al* 2025 Magnetic particle imaging angiography of the femoral artery in a human cadaveric perfusion model *Commun. Med.* **5**(1) doi:10.1038/s43856-025-00794-x

Herz S *et al* 2018 Magnetic particle imaging guided real-time percutaneous transluminal angioplasty in a phantom model *Cardiovasc. Intervent. Radiol.* **41**(7) 1100–1105 doi:10.1007/s00270-018-1955-7

Herz S *et al* 2019 Magnetic particle imaging-guided stenting *J. Endovasc. Ther.* **26**(4) 512–519 doi:10.1177/1526602819851202

Janssen K-J *et al* 2021 Single harmonic based narrowband magnetic particle imaging arXiv:2105.08786

Knopp T and Buzug T M 2012 *Magnetic Particle Imaging: An Introduction to Imaging Principles and Scanner Instrumentation* (Springer, Berlin)

Ludewig P *et al* 2022 Magnetic particle imaging for assessment of cerebral perfusion and ischemia *WIREs Nanomed. Nanobiotechnol.* **14**(1) doi:10.1002/wnan.1757

Panagiotopoulos N *et al* 2015 Magnetic particle imaging: Current developments and future directions *Int. J. Nanomedicine* **10** 3097–3114 doi:10.2147/IJN.S70488

Rahmer J *et al* 2009 Signal encoding in magnetic particle imaging: Properties of the system function *BMC Med. Imaging* **9** 1–21 doi:10.1186/1471-2342-9-4

Rahmer J *et al* 2017 Interactive magnetic catheter steering with 3-D real-time feedback using multi-color magnetic particle imaging *IEEE Trans. Med. Imaging* **36**(7) 1449–1456 doi:10.1109/TMI.2017.2679099

Rezaei B *et al* 2024 Magnetic nanoparticles for magnetic particle imaging (MPI): design and applications *Nanoscale* doi:10.1039/d4nr01195c

Salamon J *et al* Magnetic particle/magnetic resonance imaging: In vitro MPI-guided real-time catheter tracking and 4D angioplasty *PLoS One* **11**(6) doi:10.1371/journal.pone.0156899

Shen Y *et al* 2023 An adaptive multi-frame parallel iterative method for accelerating real-time magnetic particle imaging reconstruction *Phys. Med. Biol.* **68**(24) doi:10.1088/1361-6560/ad078d

Shi G *et al* 2025 Content-aware distillation network for real-time magnetic particle imaging *IEEE Trans. Instrum. Meas.* **74** doi:10.1109/TIM.2025.3535575

Sun Z *et al* 2025 Multidimensional information fusion learning for real-time signal denoising in magnetic particle imaging *IEEE Trans. Instrum. Meas.* **74** doi:10.1109/TIM.2025.3580831

Them K *et al* 2016 Increasing the sensitivity for stem cell monitoring in system-function based magnetic particle imaging *Phys. Med. Biol.* **61**(9) 3279–3290 doi:10.1088/0031-9155/61/9/3279

Top C B *et al* 2019 Trajectory analysis for field free line magnetic particle imaging *Med. Phys.* **46**(4) 1592–1607 doi:10.1002/mp.13411

Vaswani A *et al* 2017 Attention is all you need *Adv. Neural Inf. Process. Syst.* **30** 5999–6009

Vogel P *et al* 2023 iMPI: Portable human-sized magnetic particle imaging scanner for real-time endovascular interventions *Sci. Rep.* **13**(1) doi:10.1038/s41598-023-37351-2

Weaver J B *et al* 2008 Frequency distribution of nanoparticle magnetization in static and harmonic magnetic fields *Med. Phys.* **35**(5) 1988–1994 doi:10.1118/1.2903449

Weizenecker J *et al* 2009 Three-dimensional real-time in vivo magnetic particle imaging *Phys. Med. Biol.* **54**(5) L1–L10 doi:10.1088/0031-9155/54/5/L01

Wegner F *et al* 2024 Bare-metal stent tracking with magnetic particle imaging *Int. J. Nanomedicine* **19** 2137–2148 doi:10.2147/IJN.S447823

Wen J *et al* 2024 Dual-channel end-to-end network with prior knowledge embedding for improving spatial resolution of magnetic particle imaging *Comput. Biol. Med.* **178** doi:10.1016/j.compbiomed.2024.108783

Werner F *et al* 2016 Geometry planning and image registration in magnetic particle imaging using bimodal fiducial markers *Med. Phys.* **43**(6) 2884–2893 doi:10.1118/1.4948998

Zhao J *et al* 2024 MPIGAN: An end-to-end deep generative framework for high-resolution magnetic particle imaging reconstruction *Med. Phys.* **51**(8) 5492–5509 doi:10.1002/mp.17104

Zheng B *et al* 2015 Magnetic particle imaging tracks the long-term fate of in vivo neural cell implants *Sci. Rep.* **5** doi:10.1038/srep14055